\documentstyle[12pt,aasms4]{article}

\def\lsim{\lower.5ex\hbox{$\; \buildrel < \over \sim \;$}}
\def\gsim{\lower.5ex\hbox{$\; \buildrel > \over \sim \;$}}

\begin{document}

\noindent
{\it Conference Highlights}

\begin{center}

\title{\large \bf LiBeB, Cosmic Rays and Gamma-Ray Line
Astronomy\footnote{Conference was held in Paris, France in December 1998.
Proceedings will be edited by R. Ramaty, E. Vangioni-Flam, M. Cass\'e,
\& K. Olive and published in the {\it ASP Conference Series}}}

\end{center}

\medskip

\medskip

\noindent
{\it Reuven Ramaty$^1$, Elisabeth Vangioni-Flam$^2$,
Michel Cass\'e $^2$ and Keith Olive$^3$}

\noindent
{\it $^1$NASA Goddard Space Flight Center}
\vskip -0.5truecm

\noindent
{\it $^2$Institut d'Astrophysique, Paris}
\vskip -0.5truecm

\noindent
{\it $^3$University of Minnesota}

The light elements Li, Be and B (LiBeB) play a unique role in
astrophysics. The Li abundance of old halo stars is a key
diagnostic of Big Bang nucleosynthesis (BBN), along with $^2$H and
$^4$He. The essentially constant Li abundance (Li/H$\simeq 2
\times 10^{-10}$, Spite plateau) as a function of metallicity
[Fe/H] for low metallicity stars ([Fe/H]$<$-1) is believed to be
the primordial abundance resulting from BBN.
[Fe/H]$\equiv$log(Fe/H)$-$log(Fe/H)$_\odot$, where Fe/H is the Fe
abundance by number relative to H and (Fe/H)$_\odot$ is the solar
system value. 

The rare and fragile LiBeB nuclei are not generated in the normal 
course of stellar nucleosynthesis and are, in fact, destroyed in 
stellar interiors, a characteristic that is reflected in their very 
low abundances. Cosmic-ray interactions contribute to their 
production, but only $^6$Li, $^9$Be  and $^{10}$B are 
entirely cosmic-ray produced. Neutrino induced spallation, 
$^{12}$C($\nu,\nu^\prime$p)$^{11}$B appears to play an important 
role in the origin of B by producing the excess $^{11}$B needed to 
account for the B isotopic ratio in meteorites which exceeds the 
predictions of all viable cosmic-ray scenarios. While reactions on 
metals (primarily C and O) contribute to all of the LiBeB nuclei, 
reactions of fast $\alpha$ particles on ambient He produce both 
$^7$Li and $^6$Li, and are the dominant source of the latter. 
Nucleosynthesis in a variety of other Galactic objects, including 
Type II supernovae, novae and giant stars produce the bulk of the 
$^7$Li at epochs when [Fe/H] exceeds about -1. 

Traditionally, the cosmic-ray role in LiBeB evolution was 
investigated by assuming that at all epochs of Galactic evolution 
cosmic rays with energy spectra similar to those observed in the 
current epoch are accelerated out of the average interstellar medium 
and interact in the interstellar medium (ISM), mostly with C, N and 
O. This GCR paradigm, however, appears to be in conflict with 
recent  measurements of Be and B abundances in low metallicity halo 
stars, achieved with the 10 meter KECK telescope and the Hubble 
Space telescope. The GCR paradigm predicts a quadratic correlation 
of Be and B vs. Fe, as opposed to the data which show a quasi linear 
correlation. As a consequence, the paradigm has been modified 
(Cass\'e et al. 1995, Nature, 373, 318; Ramaty et al. 1996, ApJ, 
456, 525) by augmenting the cosmic rays accelerated out of the 
average ISM with a metal enriched component confined predominantly 
to low energies (\lsim 100 MeV/nucleon) and thought to be 
accelerated out of the winds of Wolf-Rayet stars and the ejecta of 
supernovae. More recently, it was suggested (Lingenfelter et al. 
1998, ApJ, 500, L153; Higdon et al. 1998, ApJ, 509, L33) that the 
cosmic rays themselves are accelerated mostly out of supernova 
ejecta rather than the average ISM, implying that the source 
material of the cosmic rays would be metal enriched at all epochs of 
Galactic evolution. Both of these models now converge towards 
acceleration by shocks in superbubbles, but they differ in the 
employed particle energy spectra, a distinction that could be tested 
by nuclear gamma-ray line observations. However, the effect is only 
marginally detectable by present generation gamma-ray telescopes.

Light element research thus impacts several important astrophysical 
problems, specifically BBN, the origin of cosmic rays, Galactic 
chemical evolution, and gamma-ray astronomy. These were then the 
topics of the Conference and they will be covered in detail in the 
upcoming Proceedings. Here we summarize some of the highlights.

Critical considerations of the flatness of the Spite plateau were 
presented by Paolo Molaro. These are essential for establishing the 
primordial Li abundance. An important issue in this context is the 
amount of Li destruction (if any) in the observed stars. Marc 
Pinsonneault and Sylvie Vauclair addressed this problem. Another 
venue for establishing the primordial nature of Li in connection 
with binaries was discussed by Francois Spite. The relationship of 
the light element data to BBN was reviewed by Keith Olive.

The $^6$Li observations in low metallicity stars were reviewed by
Lewis Hobbs. There are now good indications that $^6$Li is present
in such stars, with an abundance of a few percent relative to $^7$Li
and a factor of several tens relative to Be. The abundance ratio
relative to Be, compared with the expected ratio from the various
cosmic-ray scenarios, implies that $^6$Li could not have been
severely depleted in the stars where it is detected. Consequently,
since $^6$Li is more fragile than $^7$Li, the $^7$Li depletion
should also be small. The abundance ratio relative to $^7$Li shows
that cosmic-ray interactions could not have made a significant
contribution to the Li/H of the Spite plateau. All of these
reinforce the finding that the plateau value indeed represents the
correct primordial abundance. $^6$Li so far has been detected in
only two stars. As its production history could be quite
different from that of Be (being very efficiently produced in
interactions involving only He, unlike Be which requires the
spallation of metals), future observations over a broad range of
metallicities could lead to interesting surprises.

The very important new data on O abundances in low metallicity stars 
were presented by Ramon Garcia Lopez (see Israelian et al. 1998, 
ApJ, 507, 805). Contrary to previous data, the new observations, if 
confirmed, show that O/Fe is a monotonically increasing function of 
decreasing metallicity, reaching values that exceed the solar ratio 
by a factor of $\sim$4 at [Fe/H]$ = -1.5$ and by a full order of 
magnitude at [Fe/H]$ = -3$. Some of this increase is due to the 
absence of Type Ia supernovae in the early Galaxy. The additional 
increase is not well understood, it could be due to low Fe yields 
relative to O in the first generation of core collapse supernovae, 
or possibly due to mixing effects since, as pointed out by Audouze 
and Silk (1995, ApJ, 451, L49) the ISM of the early Galaxy, being 
metal enriched by only a small number of core collapse supernovae, 
could be quite inhomogeneous. In any case, the enhanced early 
Galactic O abundance makes cosmic-ray acceleration out of the 
average ISM more efficient. This effect, coupled with the possible 
lower Fe yield per supernova, allowed Brian Fields and Keith Olive 
to show that cosmic-ray acceleration out of the average ISM, 
hitherto believed untenable, could be viable. Their model also 
implies a decrease of $^6$Li/Be as a function of increasing 
metallicity, a result which appears to be consistent with the fact 
that the early Galactic ratio mentioned above probably exceeds the 
meteoritic ratio at solar metallicity.

A critical discussion of the NLTE effects, which are essential for 
the abundance determinations, particularly that of B, was given by 
Dan Kiselman. Douglas Duncan reviewed the B observations and Dieter 
Hartmann discussed the neutrino induced processes in core collapse 
supernovae, that in particular lead to the production of $^{11}$B. 
As already mentioned, this process provides a plausible explanation 
for the excess $^{11}$B measured in meteorites. Stellar evolution, 
another very important ingredient necessary for understanding the 
implications of the light element data, was discussed by Marc 
Pinsonneault. The status of Galactic nuclear gamma-ray line 
observations, showing that the previously reported observations of 
Orion are no longer valid, was reviewed by Hans Bloemen. In the 
absence of nuclear gamma-ray data, the detection of broad soft X-ray 
lines (particularly the lines of O just below 1 keV) resulting from 
electron capture and excitation on fast ($\sim$1MeV/nucleon) ions, 
could provide independent information on the existence of low energy 
cosmic rays. This topic was discussed by Vincent Tatischeff. The 
capabilities of the gamma-ray imaging and spectroscopic mission 
INTEGRAL, to be launched soon, were discussed by Volker 
Sch\"onfelder and Bertrand Cordier.

Current epoch cosmic-ray observations of the electron capture 
radioisotope $^{59}$Ni and its decay product $^{59}$Co, with an 
instrument on the currently active ACE mission, were presented by 
Robert Binns. $^{59}$Ni decays by electron capture with a half life 
of $7.6 \times 10^4$ years. However the decay is suppressed if the 
acceleration time scale is shorter than the lifetime because the 
atom is stripped as it is accelerated (Cass\'e and Soutoul 1975, 
ApJ, 200, L75). The fact that much more $^{59}$Co than $^{59}$Ni is 
observed, suggests a delay ($\sim$10$^5$ years) between 
nucleosynthesis and acceleration. This makes it unlikely that 
supernovae accelerate their own ejecta, but still allows cosmic-ray 
acceleration from metal enriched superbubbles, as in the Higdon et 
al. model mentioned above.

Several theoretical papers on cosmic-ray origin and acceleration 
mechanisms were presented. Jean-Paul Meyer and Donald Ellison 
reviewed their previously published model (Meyer et al. 1997, ApJ, 
487, 182; Ellison et al. 1997, ApJ, 487, 197) in which the current 
epoch cosmic rays originate from an average ISM of solar composition 
and interstellar dust plays an important role in determining the 
abundances. They also discussed the shortcomings of the recently 
proposed model (Lingenfelter et al., 1998, ApJ, 500, L153) in which 
each supernova accelerates its own freshly produced refractory 
metals. Maurice Shapiro reviewed his previously proposed model based 
on the preacceleration of the cosmic rays by coronal mass ejection 
driven shocks on low mass, cool stars. Acceleration in superbubbles 
was discussed by Andrei Bykov and Etienne Parizot, who emphasized 
that the conditions in the superbubbles that would to lead to cosmic 
rays with hard energy spectra at low energies up to a cutoff at an 
energy which is still nonrelativistic. These are the low energy 
cosmic rays which have been postulated to produce the bulk of the Be 
at low metallicities (see Vangioni-Flam et al. 1996, A\&A, 468, 
199). On the other hand, as pointed out in the publication of Higdon 
et al., since these giant superbubbles are thought to fill up a 
large fraction of the ISM, they are the most likely site for the 
acceleration of the cosmic rays, which of course show no cutoff up 
to very high ultrarelativistic energies. Thus, it is still not clear 
whether the postulated Galaxy wide low energy cosmic-ray component 
exists, a question that should be resolved by future gamma-ray line 
observations.

In summary, LiBeB research indeed spans a broad range of interesting
problems that will be covered in the planned Proceedings.

\end{document}